\documentclass[reqno,12pt]{article}
\usepackage{amsmath,amsfonts,amssymb,amsthm,amstext,amscd,eucal,srcltx}
\usepackage[all]{xy}
\usepackage{epsfig,graphicx,bm}
\usepackage{epstopdf, epsf}
\usepackage{dcolumn}
\usepackage{hyperref}
\usepackage{enumerate}

 \newcommand{\be}{\begin{equation}}
 \newcommand{\ee}{\end{equation}}
 \newcommand{\bea}{\begin{eqnarray}}
 \newcommand{\eea}{\end{eqnarray}}

\def\ps{$\boldsymbol{{\cal P}}[\chi; p_i]$ }

\newcommand{\bo}[1]{\boldsymbol{#1}}

\newcommand{\bse}{\begin{subequations}}
\newcommand{\ese}{\end{subequations}}

\begin{document}

\begin{titlepage}
\thispagestyle{empty}
   \begin{flushright}
{
}\end{flushright}
\begin{center}
    \font\titlerm=cmr10 scaled\magstep3
    \font\titlei=cmmi10 scaled\magstep4
    \font\titleis=cmmi7 scaled\magstep4
     \centerline{\titlerm Residual Diffeomorphisms and Symplectic Softs Hairs:}
\vskip2mm     
     \centerline{\large{The Need to Refine Strict Statement of Equivalence Principle}}

 \vspace{1.0cm}
    \noindent{\textbf{\large{ M. M. Sheikh-Jabbari}}}\\
    \vspace{0.8cm}

{\small\it  School of Physics,
 Institute for research in fundamental sciences (IPM),\\ P.O.Box
 19395-5531, Tehran, Iran},
\vskip 2mm
 E-mail:   jabbari@theory.ipm.ac.ir

\today
  \end{center}

  \vskip 2cm

  \begin{abstract}
General covariance is the cornerstone of Einstein's General Relativity (GR) and implies that any two metrics  related by diffeomorphisms are physically equivalent. There are, however, many examples pointing to the fact that this strict statement of general covariance needs refinement. There are a very special (measure-zero) subset of diffeomorphisms, \emph{the residual diffeomrphisms}, to which one can associate well-defined conserved charges.  This would hence render these diffeomorphic geometries physically distinct. We discuss that these symmetries may be appropriately called ``symplectic symmetries''. Existence of residual diffeomorphisms and sympelctic symmetries can be a quite general feature and not limited to the examples discussed so far in the literature.  We propose that, in the context of black holes, these diffeomorphic, but distinct, geometries may be viewed as ``symplectic soft hair'' on black holes. We comment on how this may remedy black hole microstate problem, which in this context are dubbed as ``horizon fluffs''.

\end{abstract}

\begin{center}
\vskip1cm
{\textit{Essay received Honorable Mention\\ in the Gravity Research Foundation 2016 Awards for Essays on Gravitation.}}

\end{center}
\end{titlepage}


It is well known that equivalence principle was Einstein's guide into the formulation of General Relativity (GR). Equivalence principle implies that all physical information should be available to all observers which are in causal contact with, i.e. can send and receive signals from, those events. Of course,  to make the equivalence principle precise we usually limit it to ``local observables,'' which are associated with events local in spacetime. The precise meaning of ``local event'' may be stated in a case-by-case basis. Also, in any standard Einstein GR course we are taught that (e.g. see \cite{Padmanabhan-book}) there is a coordinate system associated with any observer. Equivalence principle then implies general covariance of the theory and that physical observables should necessarily be invariant under general coordinate transformations.\footnote{As it is well known, equivalence principle is more than just general covariance and implies the minimal coupling.} 

Einstein's GR is best formulated in the mathematical framework of differential and manifold geometry,
where general covariance finds a precise statement: Dynamical degrees of freedom and field equations should be written in a covariant form in terms of tensors, which have a well defined transformation under general coordinate transformations, the \emph{diffeomorphisms}. Physical observables should then be built from scalars. Instead of field equations, we may use an action and the variational principle, where the action is a diffeomorphism invariant quantity and is given through  integral of the Lagrangian over the spacetime manifold; the Lagrangian is a scalar function of the fields. In this setup, all physical observables should hence be made from ``geometric'' objects and should be diffeomorphism invariant. In particular, consider a geometry which may be specified by a metric written in a specific  coordinate system, plus all possible coordinate extensions. ``Standard general covariance'' then implies that any metric which is related to this metric upon a coordinate transformation is physically equivalent to it. In other words, standard general covariance implies that any two metrics which are diffeomorphic to each other are physically equivalent. 

The above statement is of course a generic one in any gauge field theory, any field theory with local (gauge) symmetry: Diffeomorphisms may be viewed as gauge transformations; all tensor fields (including scalars) have a prescribed transformation under diffeomprhisms and in the terminology of gauge theories any field is ``charged'' under diffeomorphisms. Metric, as a two-tensor $g_{\mu\nu}$, is no exception and has a particular transformation:
$$
x^\mu\to x^\mu-\xi^\mu(x),\qquad g_{\mu\nu}(x)\to g_{\mu\nu}(x)+\delta g_{\mu\nu}(x),\ \delta g_{\mu\nu}= \nabla_\mu\xi_\nu+\nabla_\nu\xi_\mu.
$$
The two metrics $g$ and $g+\delta g$ are physically equivalent. Physical observables in any gauge theory are then ``gauge invariant'' quantities. For a metric these are the \emph{geometric} quantities which are independent of how we parametrize our spacetime and the coordinate system we choose. In GR, these information are the ones which could be probed or measured by different geodesics, like geodesic distance between two events (points in the spacetime), causal structure or global Killing vector fields of the spacetime. 

In the usual treatment of local gauge theories, gauge symmetries are often told to be ``redundancies of description''.  Meaning that to write the Lagrangian of the theory in a covariant way, we usually introduce extra, unphysical, gauge degrees of freedom. Gauge invariance then guarantees that these extra degrees of freedom are not local  propagating degrees of freedom. Moreover, constructing observables from  gauge invariant quantities guarantees that these extra degrees of freedom do not appear in physical observables. At the technical level, to make computations in a local (quantum)  gauge field theory, we need to fix a gauge. Then Ward-Takahashi identities guarantee that physical observables are independent of the gauge-fixing choice. In the terminology of gauge field theories, any choice of coordinate system is like fixing a gauge.
\paragraph{Gauge symmetries and conserved charges.} Symmetries and the Noether theorems have played a pivotal role in the development of our modern physics.  Symmetries subject to Noether theorems are either global-continuous ones (subject of Noether's first theorem), or  local ones (subject to Noether's second theorem). The former is related to conserved charges, while the latter leads to identities which should be satisfied to guarantee gauge invariance. These identities appear as the integrability condition of field equations at classical level and as Ward identities at quantum level. Nonetheless, one may still wonder if it is possible to associate conserved charges to local gauge symmetries. This question has been analyzed in some detail in the literature of gauge theories in general, e.g. see \cite{Soft-photon}, and differomophism invariant theories in particular. The most notable and prime examples are Bondi-Metzner-Sachs (BMS) \cite{BMS}, and Brown and Henneaux \cite{Brown-Henneaux}. As we will review below, the answer is affirmative. Existence of  states carrying these conserved charges are what we would be focusing on here.

For the purpose of this essay let us focus on $d$ dimensional gravity  theories. We can fix diffeomorphisms by choosing a coordinate system. Although our arguments are more general, to be specific, let us fix time-like Gaussian coordinates. With this choice we can fix the form of metric to
$$
ds^2=-dt^2+g_{ij}(t,x^l)dx^i dx^j, \qquad i,j=1,\cdots, d-1.
$$
We can always do so locally by appropriate choice of $d$ functions in $\xi^\mu(t,x^i)$. However, as is well known, even after this choice, the diffeomorphism freedom is not completely fixed: one has the diffeomoephisms which only depend on $x^i$. This latter differomophisms are in fact necessary for consistently removing the momenta conjugate to $g_{t\mu}$ components of metric and the associated constraints \cite{Weinberg-Cosmology}. Even after using this part of diffeomorphisms we remain with a part of diffeomorphisms which is generically  a part of, or of the same cardinality as, $d-2$ dimensional diffeomorphisms.  This remaining part is of course a measure-zero subset of original gauge freedom we started with and is not capable of removing any further propagating degrees of freedom. It is not difficult to show that these remaining diffeomorphisms together with the Lie bracket close onto an algebra which in $d\geq 3$ dimensions is infinite dimensional; it has infinitely many elements in it. If one can consistently associate well defined conserved charges to the remaining diffeomorphisms, they will not represent ``gauge,'' unphysical degrees of freedom; they would then become physical. 

\paragraph{Covariant Phase Space Method.} To examine whether there is a consistent way of associating conserved charges with the remaining diffeomorphisms, there is a powerful technical tool introduced primarily in 1987 in \cite{early-CPSM} and expanded and developed by  Robert Wald and collaborators since the earliy 1990's \cite{Lee-Wald} and enhanced and made more precise in particular by the later works of the group in ULB in Brussels, e.g. in \cite{Barnich-Brandt}. Here we only sketch the argument and the results, the technical details may be found  e.g. in \cite{Seraj-Hajian} and references therein.  Consider a solution to a generic diffeomorphism invariant gravity theory, specified by $g_{\mu\nu}$ and other fields collectively denoted by $\Phi$. This set of solutions can be specified by a set of parameters $p_i$ and are generically given in a convenient  coordinate system. Make the gauge fixings mentioned above and let $\chi$ denote the vector field generating the remaining diffeomorphisms. One may then construct continuous set of metrics generated by the successive action of the $\chi$ field on this class of solutions. At infinitesimal level, these are $g_{\mu\nu}+{\cal L}_\chi g_{\mu\nu}$, where ${\cal L}_\chi g_{\mu\nu}$ denotes the Lie derivative along $\chi$. One may do this in an active way (keeping the coordinate system and changing the components of metric tensor $g_{\mu\nu}$). We will denote this class of metrics which generically have some number of independent functions and parameters in them by $\boldsymbol{{\cal P}}[\chi; p_i]$. Constructed in this way, it is obvious that all these geometries are solutions to the original theory. These solutions for a given value of parameters $p_i$ are, however, deemed to be physically equivalent according to the ``standard general covariance.''

\paragraph{Solution Phase Space.} The Covariant Phase Space Method (CSPM) may then be used to promote $\boldsymbol{{\cal P}}[\chi; p_i]$ to a phase space, the \emph{solution phase space}, by providing a symplectic structure current: $\boldsymbol{\omega}(\delta_1\Phi, \delta_2\Phi; \Phi)$, $\boldsymbol{\omega}$ is a $(d-1)$-form on the spacetime while a two-form on the phase space $\boldsymbol{{\cal P}}[\chi; p_i]$ and $\delta\Phi$ are generic elements in the tangent space of $\boldsymbol{{\cal P}}[\chi; p_i]$. Since $d\boldsymbol{\omega}\approx 0$ ($\approx$ means on-shell equality, i.e., $\delta\Phi$ are satisfying linearized field equations and $\Phi$ is a solution), then  $\boldsymbol{\omega}\approx d\boldsymbol k$ and in particular, $\boldsymbol{\omega}(\delta\Phi, \delta_\chi\Phi; \Phi)\approx d\boldsymbol{k}_\chi[\delta\Phi;\Phi]$, where $\bo{k}_\chi$ is a $(d-2)$-form on spacetime and a one-form on $\boldsymbol{{\cal P}}[\chi; p_i]$. Charge variation associated with the field variations $\delta_\chi\Phi$ is defined as a surface integral, an integral over a codimension two spacelike compact surface $\Sigma$
$$
\delta Q_\chi=\oint_{\Sigma} \boldsymbol{k}_\chi[\delta\Phi;\Phi].
$$
If the ``integrability condition'' is met (e.g. see \cite{Lee-Wald, Seraj-Hajian, Hajian-me}) then $\delta Q_\chi$ is an exact one-form on $\boldsymbol{{\cal P}}[\chi; p_i]$ and one can hence  define the conserved charge $Q_\chi$ by integrating charge variation over an arbitrary path in the phase space. These charges are also called ``Hamiltonian generators'' of the $\chi$-transformations. 

Given a vector field $\chi$ we may, in principle, compute $\delta Q_\chi$. If $\delta Q_\chi$ are vanishing, $\chi$'s generate ``pure gauge transformations'' on the phase space. One needs to mod out the phase space $\boldsymbol{{\cal P}}[\chi; p_i]$ with these pure gauges to define the physical phase space \cite{Lee-Wald}. We will define our \emph{residual diffeomorphism} after modding out the set of $\chi$'s for which $\delta Q_\chi$ are finite, by such pure gauge diffeomorphisms.

\paragraph{Symplectic symmetries and charges.} It may happen that, as has been demonstrated in some different examples \cite{Hajian-me}, for a set of $\chi$'s the presymplectic form $\bo{\omega}$ vanishes on-shell:
$$ 
\bo{\omega}(\delta_\chi\Phi, \delta\Phi;\Phi)\approx 0.
$$
In this case one may readily prove that the charge $\delta Q_\chi$ is conserved and is independent of $\Sigma$, which may now be chosen arbitrarily. In this case we are dealing with \emph{symplectic symmetries}. Based on different examples discussed e.g. in \cite{Seraj-Hajian}, I will assume that the set of the residual diffeomorphisms are not empty and that they are always related to symplectic symmetries. Symplectic symmetries may come into two families \cite{Hajian-me}, when $\chi$ generates an exact symmetry of the solution, which will be denoted by $\eta$, with $\delta_\eta\Phi=0$, or when $\delta_\chi\Phi\neq 0$. The Killing vectors are a part of the set of $\eta$'s.

To each of these two sets one can associate \emph{symplectic charges}. For the symplectic non-exact symmetries, the charge is defined as we discussed above. For the exact symmetries, we define the charges for \emph{parametric variations}, $\delta_{p_i}\Phi$, explicitly,
$$
\delta_{p_i}Q_\eta=\oint_{\Sigma} \boldsymbol{k}_{\eta}[\delta_{p_i}\Phi;\Phi],\qquad \bo{\omega}(\delta_{p_i}\Phi, \delta_\eta\Phi;\Phi)\approx d \boldsymbol{k}_{\eta}[\delta_{p_i}\Phi;\Phi].
$$
The set of $Q_{\eta}$ are essentially the ADM charges for asymptotically flat solutions and also include  Wald's entropy \cite{Wald-entropy}. 
Since the set of $\chi$'s are generically defined based on arbitrary functions, their number is infinite (while usually countable).  The $(Q_{\eta}, Q_\chi)$ charges may be used to specify points on the solution phase space $\boldsymbol{{\cal P}}[\chi; p_i]$. By construction, we expect this labelling to be unique up to possibly some ``topological'' discrete charges. 

From a different viewpoint, built upon the symmetries of symplectic structure, the symplectic charges $Q_{\eta}, Q_\chi$ may be viewed as the generators of symmetry directions on the phase space. As generators of symmetries, one may study the algebra of these charges. A fundamental theorem  in the CPSM states that, e.g. see \cite{Seraj-Hajian} and references therein, 
$$
[Q_{\chi}, Q_{\tilde{\chi}}]=Q_{[\chi,\tilde{\chi}]}+ C_{(\chi,\tilde{\chi})},\qquad [Q_\chi, Q_{\eta}]=0,
$$
where $C$ is a central element of the algebra (which may in principle be a function of $Q_\eta$).

\paragraph{Residual diffeomorphisms and ``symplectic soft hair'' on black holes.}\footnote{The expression ``soft hair'' was coined in \cite{Hawking} and denotes states which are charged under residual diffeomorphisms.} As we argued elements in the solution phase space \ps are to be viewed as physically distinct solutions, while as far as the ``standard general covariance'' is concerned only elements with distinct $p_i$ are distinguishable; classical gravity observers are blind to the $Q_\chi$ charges and can only resolve $Q_\eta$'s. 

One may try to apply the above general arguments and picture to black holes. In this setting our proposal is very simply stated as: for a black hole solution with parameters $p_i$, which may be (uniquely) specified by $Q_\eta$, there are ``hairs'' labelled by $Q_\chi$. The black hole microstates and degrees of freedom relevant to  (thermo)dynamics of black holes are elements of the phase space $\boldsymbol{{\cal P}}[\chi; p_i]$. In a different wording, our proposal is as follows:

\emph{To any black hole, one may associate two kinds of information: Classical local information and ``quasi-local'' semi-classical information. The classical information are the usual geometric quantities which we are familiar with from the usual GR and standard general covariance. These information are  available to classical local probes/geodesics and include geodesic distances, causal and horizon structure and usual ``thermodynamical'' quantities associated with black holes, like their mass, angular momentum, electric charge, surface gravity and horizon angular velocity and electric potentials. On the other hand, there are the symplectic charges $Q_\chi$. These charges are semi-classical and non-local as they cannot be measured by classical local observers, they are given by surface integrals. Being symplectic charges, they could be defined on generic  compact, codimension two spacelike surface $\Sigma$. Geometries distinguished by these symplectic charges, the ``symplectic soft hair,'' are all diffeomorphic to each other and are hence deemed the same from a usual classic GR viewpoint. A black hole and its symplectic hair form a phase space $\boldsymbol{{\cal P}}[\chi;p_i]$. The above clearly states the necessity to revise the strict statement of equivalence principle and general covariance discussed in the opening of this essay, so that it includes and accommodates the residual symmetries and the associated charges. 
}

Our vision and hope is that these ideas are relevant to identification of black hole microstates as states carrying non-trivial symplectic symmetry charges. The first examples towards explicit realization of this vision and hope has been taken in \cite{NH-soft-hair, AGS, Hossein-Shahin}. Here we sketch the ``horizon fluffs'' or ``fluff ball'' proposal put forward in \cite{AGS} and for the details of technicalities the interested reader may look at these papers and references therein:

\emph{In geometries with (event) horizon like black holes, one has the possibility of distinct residual symmetries in the near horizon  and asymptotic regions. In particular, a black hole state is the one identified with asymptotic symmetries (which include ADM-type charges like mass and angular momenta). Moreover, there are near horizon residual symmetries which may not be extended to the asymptotic region. The fluff ball proposal then states that the black hole microstates, horizon fluffs, are the states which are distinguishable by the near horizon residual symmetry charges. The horizon fluffs of a given black hole are, however, indistinguishable by their asymptotic charges.}

The above proposal has been worked through in \cite{AGS, Hossein-Shahin} for generic AdS$_3$ black holes, while is expected to be generalizable to more realistic astrophysical black holes which may be approximated by Kerr geometries. Our ``fluff ball'' proposal has some similarities with the ``fuzz ball'' proposal of Samir Mathur \cite{Fuzzball}, while it has some basic differences with it. In the fuzz ball proposal  microstates correspond to an ensemble of geometries, each of which is smooth and horizon free, and are classically distinct within the usual classical GR. Its realization has been successful in supersymmetric D1D5-P setting in string theory and relies on supersymmetry and string theory machinery, e.g. see \cite{Bena} and references therein. In the fluff ball proposal, however, the microstates are all geometries which are diffeomorphic to a given near horizon geometry (and not distinct in the standard general relativity sense); they differ in soft hairs. That is, they all have the same near horizon and causal structure, they all correspond to the same black hole with the specified asymptotic charges.

Here we discussed  surface charges associated with residual diffeomorphisms and that their presence calls for revisiting strict notion of equivalence principle and general covariance. As a possible application we used existence of states labelled by certain residual diffeomorphisms to identify black hole microstates, which in this context we called them horizon fluffs. One may then wonder if these ``soft hairs'' can be used to discuss more dynamical questions regarding black holes, questions like information paradox. Some early steps in this direction has been taken in \cite{Hawking, Compere}.  Moreover, for the asymptotic AdS geometries where we have a dual CFT picture, one may wonder if there is any relation between our proposal and the CFT description. Some early discussions on this in the AdS$_3$ example was presented in \cite{AdS3}, where it was argued that the ``symplectic soft hair'' are indeed fully capturing the so-called ``boundary gravitons,'' the degrees of freedom of presumed dual 2d CFT. We hope our proposal here opens a new window onto the celebrated AdS/CFT correspondence \cite{AdS/CFT}.


\paragraph{Acknowledgements.} I would like to thank my collaborators Hamid Afshar, Geoffrey Comp\'ere, Daniel Grumiller, Kamal Hajian, Ali Seraj, Joan Simon and Hossein Yavartanoo for many discussions over the years and for their role in the development of the  ideas presented in this essay. I would also like to thank Glenn Barnich, Mehrdad Mirbabayi, Marco Serone and Marko Simonovi\'c  for discussions or comments. This work is supported in part by  grants from ICTP NET-68, ICTP Simons fellowship, Allameh Tabatabaii grant prize of Boniad Melli Nokhbegan of Iran and by SarAmadan grant of Iranian vice presidency in science and technology. I would also like to thank ICTP for the hospitality where this note was written down.

{}

\end{document}